\documentclass[aps,twocolumn,secnumarabic,balancelastpage,amsmath,amssymb,nofootinbib,floatfix]{revtex4-1}

\usepackage{graphicx}      
\usepackage[colorlinks=true]{hyperref}  
\usepackage{array}
\usepackage{amsmath}
\usepackage{blkarray}
\usepackage{breqn}
\usepackage{enumerate}

\usepackage{subcaption}

\begin{document}

\title{Nanophotonic Particle Simulation and Inverse Design Using Artificial Neural Networks}
\author{John Peurifoy}
\author{Yichen Shen}
\email{ycshen@mit.edu}
\author{Li Jing}
\author{Yi Yang}
\author{Fidel Cano-Renteria}
\author{Brendan Delacy}
\author{Max Tegmark}
\author{John D. Joannopoulos}
\author{Marin Solja\u{c}i\'{c}}
\date{\today}
\affiliation{MIT Department of Physics}

\begin{abstract}

We propose a method to use artificial neural networks to approximate light scattering by multilayer nanoparticles. We find the network needs to be trained on only a small sampling of the data in order to approximate the simulation to high precision. Once the neural network is trained, it can simulate such optical processes orders of magnitude faster than conventional simulations. Furthermore, the trained neural network can be used solve nanophotonic inverse design problems by using back-propogation - where the gradient is analytical, not numerical. 

\end{abstract}

\maketitle


Inverse design problems are pervasive in physics \cite{scatTheory,photonicDevice,thinFilmMat,heatInverse}.  Quantum Scattering Theory \cite{scatTheory}, photonic devices \cite{photonicDevice}, and thin film photovoltaic materials \cite{thinFilmMat} are all problems that require inverse design. A typical inverse design problem requires optimization in high dimensional space, which usually involves lengthy calculations. For example, in photonics, where the forward calculations are well understood with Maxwell's equations, solving one instance of an inverse design problem can often be a substantial research project. 


There are many different ways to solve inverse design problems, which can be classified into two main categories: the genetic algorithm \cite{geneticRoy,geneticFroemming} (searching the space step by step), and adjoint method \cite{adjointGiles} (mathematically reversing the equations). For problems with many parameters, solving these with genetic algorithms takes a lot of computation power and time, and this time grows exponentially as the number of parameters increases. On the other hand, the adjoint method is far more efficient than the genetic algorithms; however, setting up the adjoint method often requires a deep knowledge in photonics, and can be quite non-trivial even with such knowledge. 

Neural Networks (NNs) have previously been used to approximate many physics simulations with high degrees of precision. Recently Carleo et. al. \cite{Carleo602} used NNs to solve many-body quantum physics problems, and Faber et. al. \cite{2017arXiv170205532F} used NNs to approximate Density Functional Theory. In this paper, we propose a novel method to further simulate light interaction with nanoscale structures and solve inverse design problems using Artificial Neural Networks. In this method, a neural network is first trained to approximate a simulation; thus the neural network is able to map the scattering function into a continuous, higher order space where the derivative can be found analytically. The "approximated" gradient of the figure of merit (FOM) with respect to input parameters is then obtained analytically with standard back-propagation \cite{hintonBack}. The parameters are then optimized efficiently with the gradient descent method. Finally, we compare our performance with the standard gradient free optimization method and find our method is orders of magnitude faster and more effective than traditional methods.




\begin{figure*}[htb]
\includegraphics[width=\linewidth]{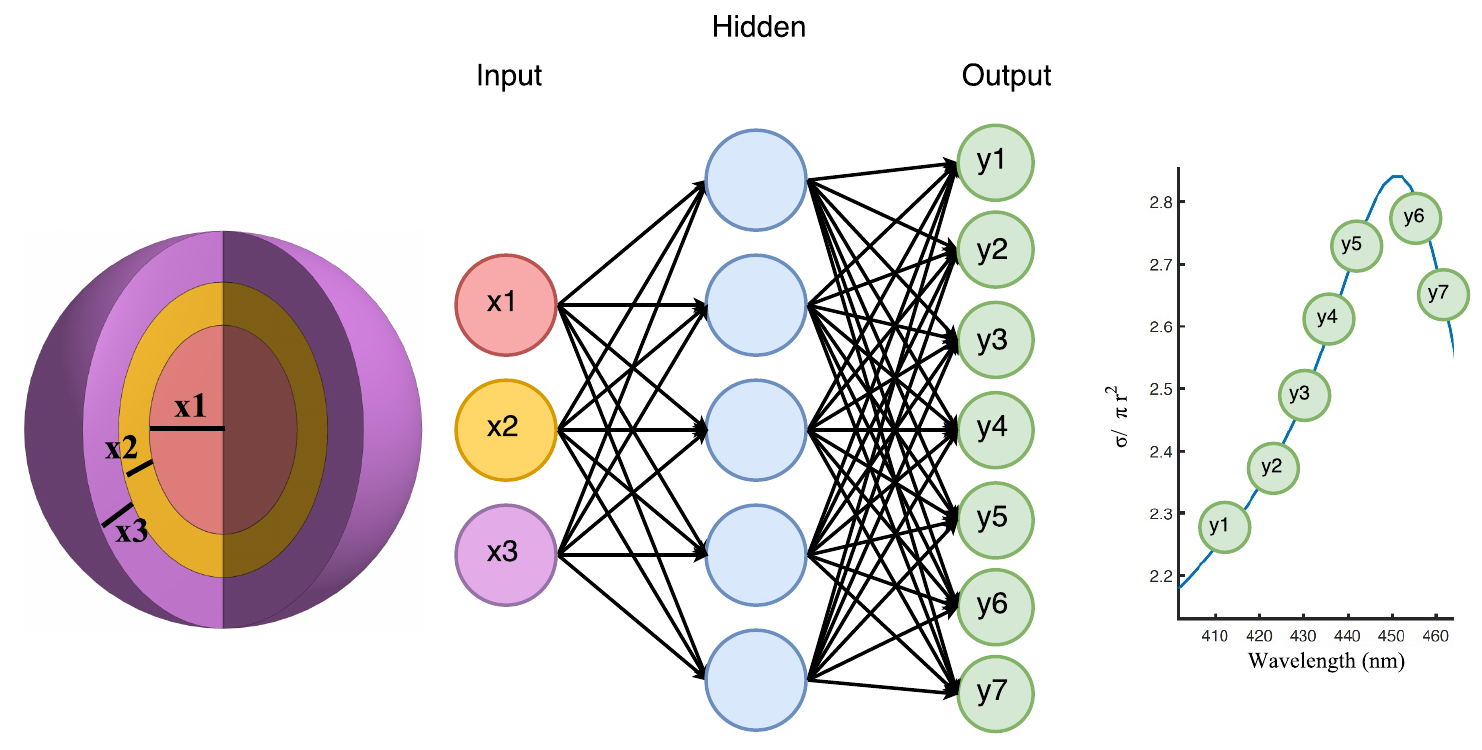}
\caption{The neural network architecture has as its inputs the thickness of each layer of the nanoparticle, and as its output the scattering cross section at different wavelengths of the scattering spectrum. Our actual neural network has four hidden layers.}
\label{fig:nn_demo}
\end{figure*}





While we focus here on a particular problem of light scattering from nanoparticles, the approach presented here can fairly easily be generalized to many other nanophotonic problems. This approach offers both the generality present in numerical optimization schemes (where only the forward calculation must be found), and the speed of an analytical solution (owing to the use of an analytical gradient). Conceptually, there are a number of reasons why the approach used here is useful for a myriad of branches of physics. After the neural network is trained, there are three key uses discussed here:

\begin{enumerate}
\item
Approximate --- Once the neural network is trained to approximate a complex physics simulation (such as density functional theory or finite difference time domain simulation), it can approximate the same computation in orders of magnitude less time. 
\item
Inverse Design --- Once trained, the neural network can solve inverse-design problems more quickly than its numerical counterpart, because the gradient can be found analytically instead of numerically. Furthermore, the series of calculations for inverse design can be computed more quickly due to the faster forward calculation. Finally, the neural network can search more easily for a global minimum because the space is smoothed in the approximation. 
\item
Optimization --- Similarly to inverse design, the network can be asked to optimize for a desired property. This functionality can be implemented simply by changing the cost function used for the design and without retraining the neural network. 
\end{enumerate}

\section{Results}

\subsection{Neural Networks can learn and approximate Maxwell Interactions}

We evaluate this method by considering the problem of light scattering from a multi-layer dielectric spherical nanoparticle --- Fig.~\ref{fig:nn_demo}. Our goal is to use a Neural Network to approximate this simulation. For definiteness, we choose a particle that has a lossless silica core ($\epsilon = 2.04$), and then alternating lossless $TiO_2$ ($\epsilon = 5.913 + \frac{.2441}{\lambda^2-.0803}$) and lossless silica layers. Specifically, we consider eight layers between 30nm to 70nm thicknesses per layer. Thus the smallest particle we consider is 480nm in diameter, and the largest is 1,120nm. 

This problem can be solved analytically or numerically with the Maxwell equations, though for multiple layers, the solution becomes involved. The analytical solution is well known \cite{WileyBook}. We used the simulation to generate 50,000 examples from these parameters with Monte-Carlo sampling.

\begin{figure*}[htb]
\begin{subfigure}{.465\linewidth}
\centering
\includegraphics[width=\linewidth]{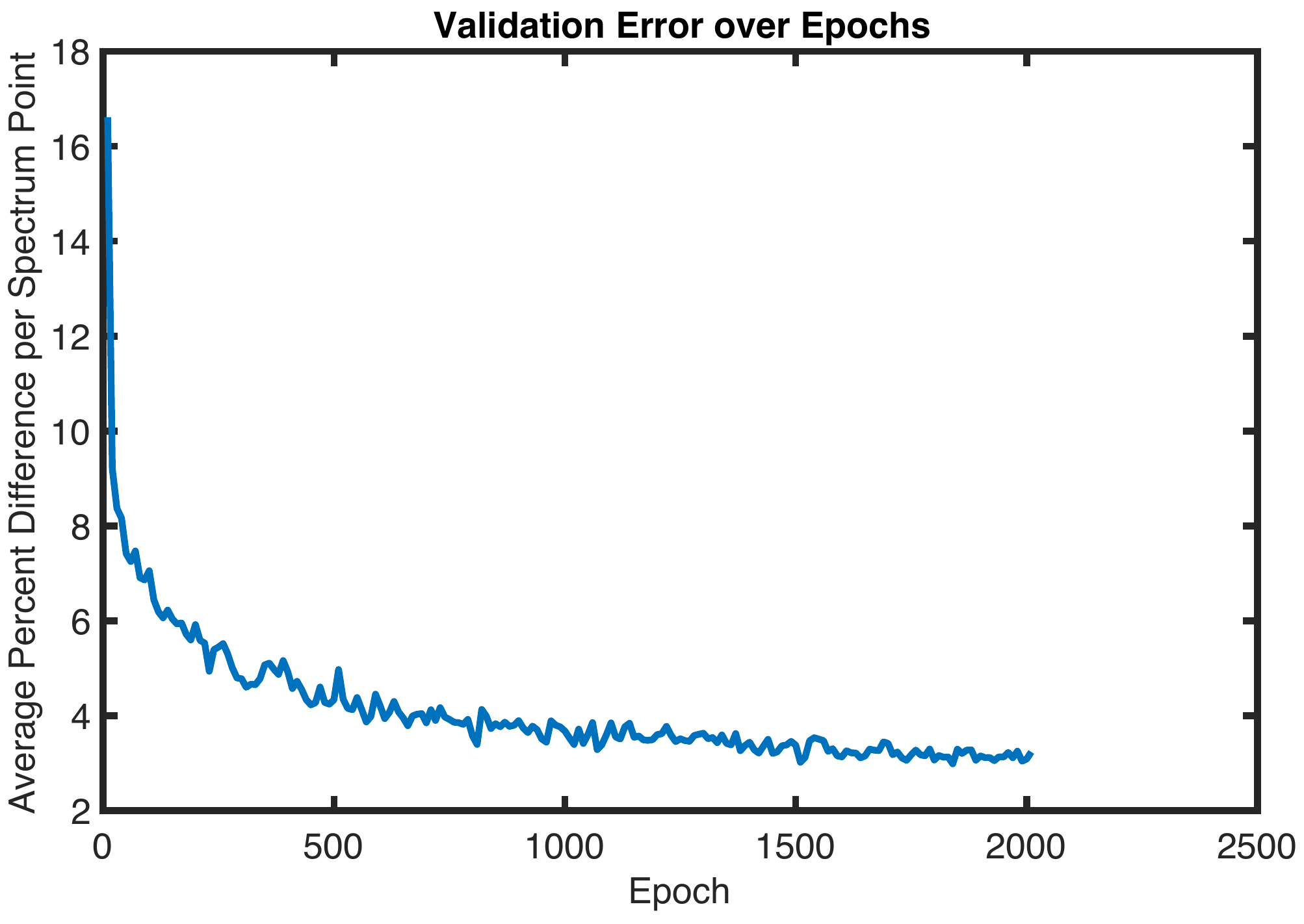}
\caption{}
\label{fig:lossFile}
\end{subfigure}
\begin{subfigure}{.52\linewidth}
  \centering
  \includegraphics[width=\linewidth]{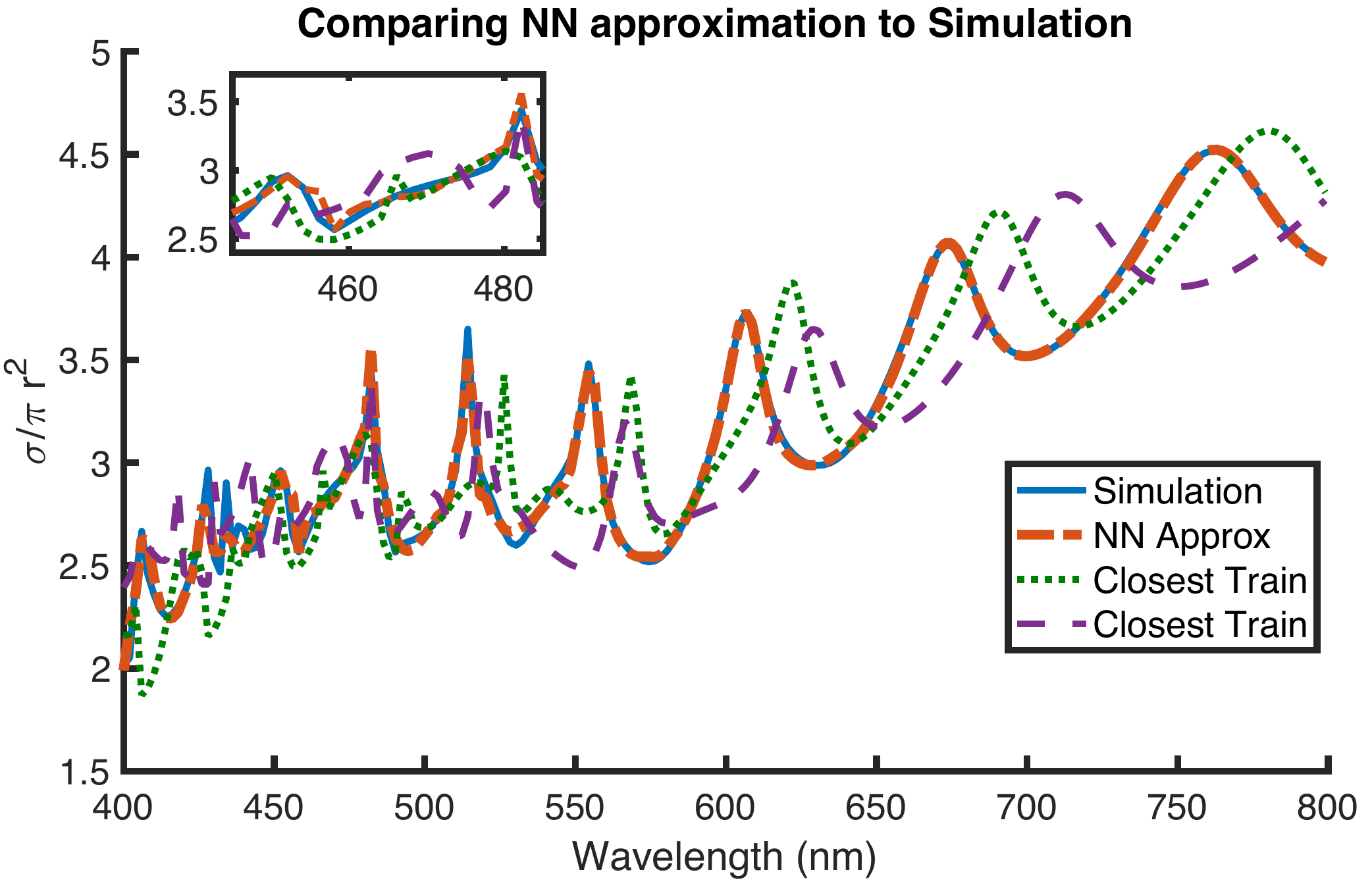}
  \caption{}
  \label{fig:Spect1Fixed}
\end{subfigure}%
\caption{a) Training loss for the eight layer case. The loss has sharp declines occasionally, suggesting that the neural network is `learning' something about the data at each point. b) Comparison of neural network approximation to the real spectrum, with the closest training examples shown here. One training example is the most similar particle larger than the desired, and the other is the most similar particle smaller than desired.  These results were consistent across many different spectra.}
\label{fig:ncnot}
\end{figure*} 

Next, we trained the neural network using these examples. We used a fully connected network, with four layers and 250 neurons per layer, giving us 239,500 total parameters. The input was the thickness of each layer (the materials were fixed), and the output was the spectrum sampled at points between 400 to 800 nanometers. The training error is graphed in Fig.~\ref{fig:lossFile}. Once the training was complete, the weights of the neural network are fixed and saved into files which can be easily loaded and used. Next, we began to experiment with applications and uses of this neural network.

The first application was to test the forward computation of the network to see how well it approximates the spectra it was not trained on --- for an example see Fig.~\ref{fig:Spect1Fixed}. Impressively, the network matches the sharp peaks and high Q features with much accuracy, even though the model was only trained with 50,000 examples --- which is equivalent to sampling each layer thickness between 30-70 nanometers only four times. 

To study if the network learned anything about the system and can produce features it was not trained on, we also graphed the closest examples in the training set. The results show that the network is able to match quite well spectra even outside of the training set. Furthermore, the results from Fig.~\ref{fig:Spect1Fixed} visually demonstrate that the network is not simply interpolating, or averaging together the closest training spectra. This suggests that the neural network is not simply fitting to the data, but instead learning some pattern about the input and output data such that it can solve problems it had not encountered, and to some extent generalize the physics of the system. 

This method is similar to the well known surrogate modeling \cite{surrogateModel}, where it creates an approximation to solve the computationally expensive problem, instead of the exact solution. However, the result indicates neural networks can be very powerful in approximating linear optical phenomena (such as nanoparticle scattering phenomena shown here). 


\subsection{Neural Networks solve Nanophontonic Inverse Design}

For inverse design, we want to be able to draw any arbitrary spectrum, and find the geometry that would most closely produce this spectrum. 

Neural networks are able to solve inverse design problems efficiently. With the weights fixed, we set the input as a trainable variable and used back-propogation to train the inputs of the neural network. In simple terms, we run the NN `backwards'. To do this, we fix the output to the desired output, and let the neural network `learn' the correct inputs. After a few iterations, the neural network suggests a geometry to reproduce the spectrum.

\begin{figure*}[htbp]
\includegraphics[width=1.0\linewidth]{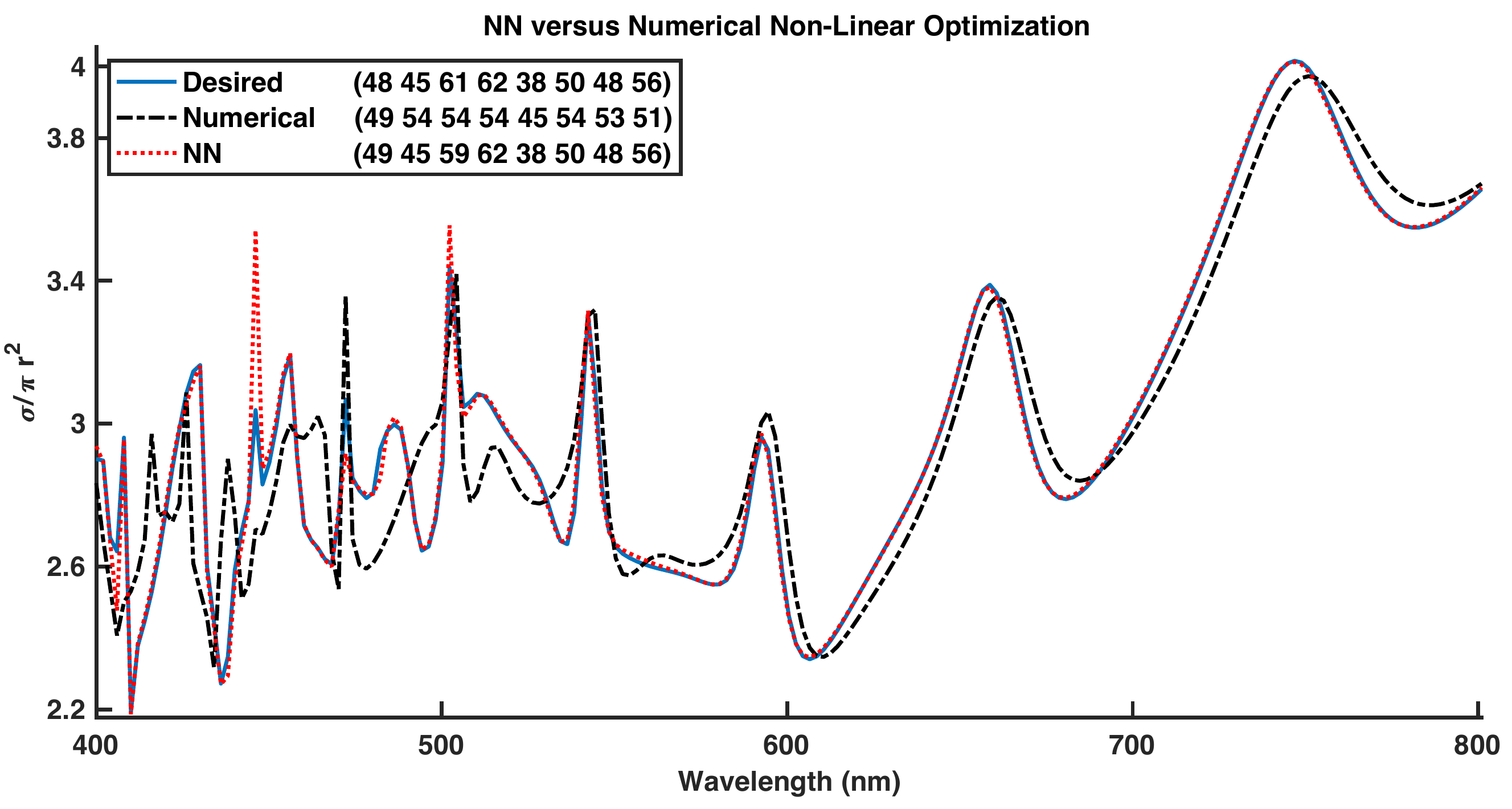}
\caption{Inverse design for an eight layer nanoparticle. The legend gives the dimensions of the particle, and the blue is the desired spectrum. The neural network is seen to solve the inverse design much more accurately.
}
\label{fig:8_layer_invert}
\end{figure*}

We test this inverse design on the same problem as above - an eight layer nanoparticle made of alternating layers of $TiO_2$ and silica.  We choose an arbitrary spectrum, and have the network learn what inputs would generate a similar spectrum. We can see an example optimization in Fig.~\ref{fig:8_layer_invert}. In order to ensure that we have a physically realizable spectra, the desired spectrum comes from a random valid nanoparticle configuration.

We also compare our method to state of the art numerical nonlinear optimization methods. We tested several techniques, and found that interior-point methods \cite{interiorPoint} were most effective for this problem. We then compared these interior-point methods to our results from the neural network, shown in Fig.~\ref{fig:8_layer_invert}. Visually, we can see that the neural network is able to find a much closer minimum than the numerical nonlinear optimization method. This result is consistent across many different spectra, as well as for particles with different number of layers and materials.

We found that for few parameters to design over (for three to five dielectric layers), the numerical solution presented a more accurate inverse design than the neural network. However, as more parameters were added (regimes of five to ten dielectric layers), the numerical solution quickly became stuck in local minima and was unable to solve the problem, while the neural network still performed well and found quite accurate solutions to inverse design. Thus, for difficult inverse design problems involving many parameters, neural networks can often solve inverse design easily. We believe this is because the optimization landscape is smoothed in the approximation.

We further studied how the Neural Network behaves in regions where $\epsilon$ has a strong dependence on $\omega$, such as the case of J-Aggregates \cite{jAgg}. This material produced complex and sharp spectra, and it is interesting to study how well the Neural Network approximated these particles, particularly for particles that it had not trained on. Results demonstrated the network was able to behave fine in these situations --- see \ref{jaggs}.

\subsection{Neural Networks can be used as an optimization tool for broadband and specific-wavelength scattering}

For optimization, we want to be able to give the boundary conditions for a model (for instance how many layers, how thick of a particle, what materials it could be), and find the optimal particle to produce $\sigma(\lambda)$ as close as possible to the desired $\sigma_{desired}(\lambda)$.

Now that we can design an arbitrary spectrum using our tool with little effort, we can further use this as an optimization tool for more difficult problems. Here, we consider two: how to maximize scattering at a single wavelength, while minimizing the rest, and how to maximize scattering across a broad-spectrum, while minimizing scattering outside of it.

To do this, we fix the weights of the neural network, and create a cost function that will produce the desired results. We simply compute the average of the $\sigma(\lambda)$ inside of the range of interest, and compute the average of the points outside the range, then minimize this ratio.This cost function $J$ is given by

\begin{equation}
J = \frac{\overline{\sigma_{in}}}{\overline{\sigma_{out}}}
\end{equation}

\begin{figure*}[htb]
\begin{subfigure}{.452\linewidth}
\centering
\includegraphics[width=\linewidth]{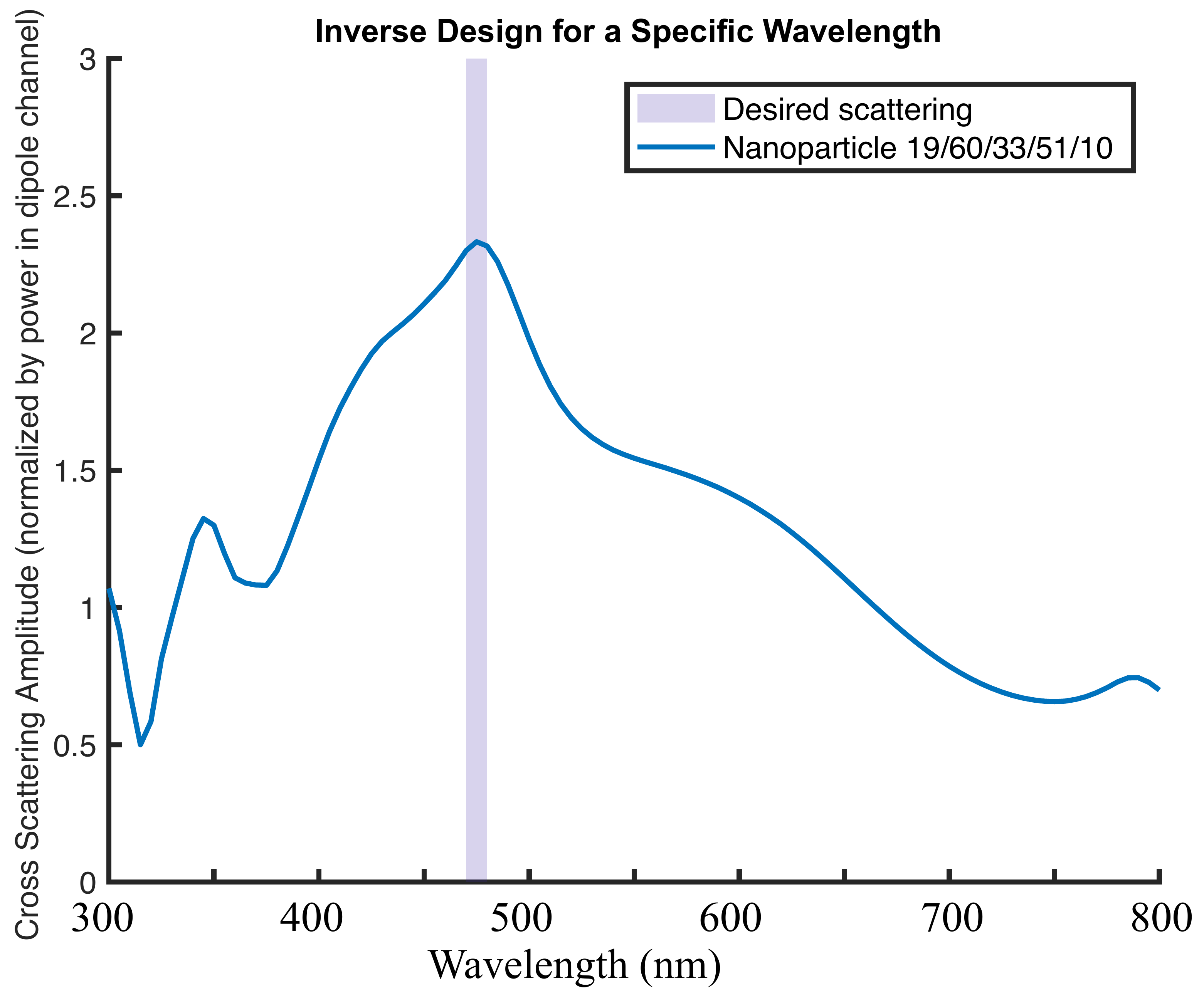}
\caption{}
\label{fig:match_a}
\end{subfigure}
\begin{subfigure}{.483\linewidth}
  \centering
  \includegraphics[width=\linewidth]{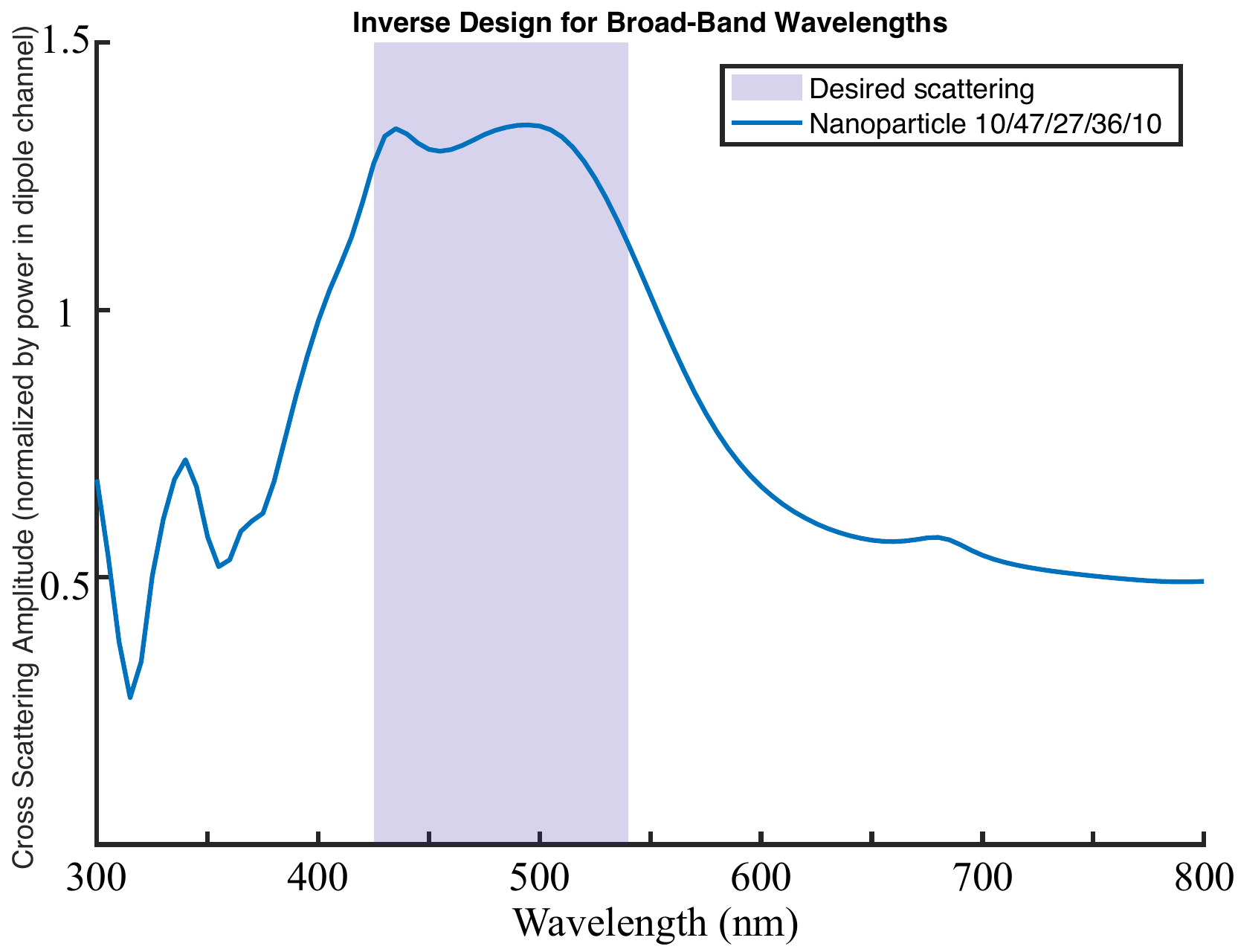}
  \caption{}
  \label{fig:match_b}
\end{subfigure}%
\caption{Spectra produced by using our approach as an optimization tool. The materials here were constrained to be solely dielectric materials, without any metals or plasmonic resonances. a) demonstrates scattering at a narrow range close to a single wavelength. Here, we force the neural network to find a total geometry that scatters around a single peak, despite the underlying materials not being able to. b) Demonstrates scattering across a broad-band of wavelengths. The legend specifies the thickness of each layer in nm, alternating TiO2 and silica layers. The network here was restricted to fewer layers of material (only five layers), but given a broader region of layer sizes than previously (from 10nm-70nm).\label{fig:match}}
\end{figure*} 

Ideally, this optimization would be performed using metals and other materials with plasmonic resonances \cite{jAgg} in the desired spectrum range. These materials are well-suited for having sharp, narrow peaks, and as such can generate spectra that are highly efficient at scattering at precisely a single wavelength. 

Our optimization here uses solely dielectric materials. By using materials that do not have sharp plasmonic resonances, we force the neural network to find a total geometry that still scatters at a single peak, despite the underlying materials being unable to. A figure showing the results of this for a narrow set of wavelengths close to 465 nanometers can be seen in Fig.~\ref{fig:match_a}.

Next, we consider the case of broadband scattering, where we want a flat spectrum across a wide array of wavelengths. In this case, we choose the same $J$ as above - minimizing the ratio of values inside to outside. After training the network for a short number of iterations, we achieve a geometry that will broad-band scatter across the desired wavelengths. A figure of this can be seen in Fig.~\ref{fig:match_b}.

\subsection{Comparison of Neural Networks with some conventional Inverse Design Algorithms}

\begin{figure*}[htbp]
  \centering
  \includegraphics[width=0.7 \linewidth]{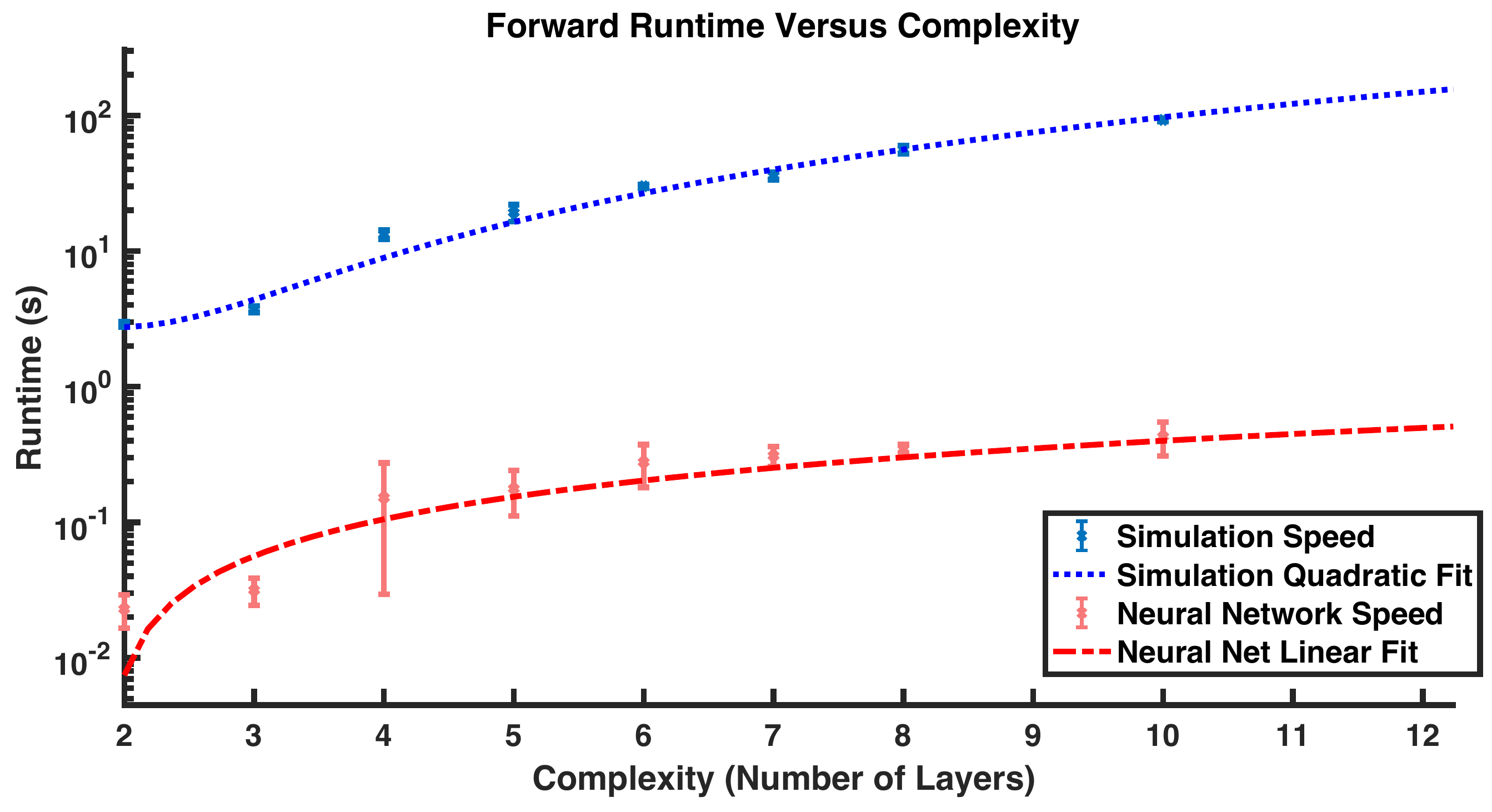}
\caption{Comparison of forward runtime versus complexity of the nanoparticle. The simulation becomes infeasible to run many times for large particles, while the neural network's time increases much more slowly. Conceptually this is logical as the neural network is using pure matrix multiplication --- and the matrices do not get much bigger --- while the simulation must approximate higher and higher orders. The scale is log-log. The simulation was fit with a quadratic fit, while the neural network was a linear fit. See \ref{inverseRuntime} for more details and inverse design speed comparison.
}
\label{fig:Forward_Rate_Comp}
\end{figure*} 

As mentioned, we tested several techniques, and found that interior-point methods \cite{interiorPoint} were most suited for nanoparticle inverse design. To compare this numerical nonlinear optimization method to our neural network, we use the same cost function for both - namely that of the mean square distance between points on the spectra. For definiteness, we code both the neural network and simulation in Matlab. This allows for reasonably fair comparisons of speed and computation resources. 

We train a different neural network on each number of particle layers from two to ten. The networks' size increased as we increased the number of layers, and the training can often require substantial time. However, once the networks were trained the runtime of these was significantly less than the forward computation time of the simulation. We tested this by running 100 spectra, then finding the average time required for the computation. These were run on a 2.9 GHz Intel Core i5 processor, and all were parrallized onto 2 CPU's. A plot of these results is shown in Fig.~\ref{fig:Forward_Rate_Comp}. Once fitting with lines, it is evident that if the problem becomes complex, the simulation would struggle to run more than a few layers, while the neural network would be able to handle more. Thus, the neural network approach has much to offer to physics and inverse design even in just speeding up and approximating simulations.

Next, we looked at the optimization runtime versus the complexity of the problem, once again comparing our method against Matlab's default algorithms. To find the speed of this optimization, we chose a spectrum and set a threshold cost, and timed how long it took for the programs to find a spectrum that is below that cost or converged into a local minimum. On a number of spectra, we found that neural networks were often sensitive to initialization points. In order to investigate these results rigorously, and not be influenced by the choice of initial conditions, we took 50 starting points for each spectrum, and tested three spectra for each number of layers. Results demonstrated that inverse design using the neural network was able to handle more complex problems than the numerical inverse design --- see \ref{inverseRuntime}.



\section{Discussion}


The results of this method suggest that it can be easily used and implemented, even for complex inverse design problems. The architecture used in the examples above --- a fully connected layer --- was chosen without much optimization, and still performs quite well. Our preliminary testing with other architectures (convolutions, dropouts, and residual networks) appeared to have further promise as well. 

Perhaps the two most surprising results were how few examples it takes for the network to approximate the simulation, as well as how complex the approximation can really be. For instance, in the eight layer case the NN only saw 50,000 examples over eight independent inputs. This means that on average it sampled only four times per layer thickness, and yet could reproduce the entire range of 30-70 nanometer layer thickness continuously. The approximation was even able to handle quite sharp features in the spectrum that it otherwise had not seen.

Promising and effective results have been seen by applying this method to other nanophotonic inverse design problems. Recently, Dianjing et. al. \cite{trainDeep} demonstrated that by using a bi-directional neural network \cite{trainBiDir}, optimization and inverse design can be performed for one dimensional layers of dielectric mediums. The approach was to first train the network to approximate the forward simulation, then do a second iteration of training to further improve the accuracy of the results. By utilizing a second iteration of training, Dianjing et. al \cite{trainDeep} was able to overcome degeneracy problems wherein the same spectrum can be generated by particles of different geometrical arrangements. Overall, this and similar work is promising to the idea that experimenting with different architectures, and adding more training data, can allow these neural networks to be useful for solving inverse design in many more scenarios. 

One clear concern with the method is that we still have to generate the data for each network, and this takes up time for each inverse design problem. It is true that generating the data takes significant effort, but there are two reasons why this method is still very useful. First, hardware is cheap, and the generation of data can be done easily in parallel across machines. This is not true for inverse design. Inverse Design must often be done in a serial approach as each step gets a little closer to the optimal, so the time cannot be reduced significantly by parallel computation. The second reason this method is highly valuable is because while the forward propagation is linear in complexity, the optimization is polynomial. Specifically, by looking at Fig.~\ref{fig:Forward_Rate_Comp} and Fig.~\ref{fig:opt_graphs_1}, we can see that the inverse design speed is growing much faster than the forward runtime. This is important because it means that for complex simulations, the numerical inverse design could take an infeasible amount of time, while the NN forward calculation may not take long; it will simply have many variables. 

This method could be used in many other fields of computational physics; it would allow us to approximate physics simulations in fractions of the time. Furthermore, owing to the robustness of back-propogation, this method allows us to solve many inverse design problems without having to manually calculate the inverse equations. Instead, we simply have to write a simulation for the forward calculation, and then train the model on it to easily solve the inverse design.

\section{Methods}

\subsection{Analytically Solving Scattering via the Transfer Matrix Method}\label{analyticalSolution}

We use the transfer matrix method, described in \cite{analyticalSolution}. We consider a multilayer nanoparticle. Due to spherical symmetry, we decompose the field into two parts: Transverse Electric (TE) and Transverse Magnetic (TM). Both these potentials satisfy the Helmholtz equation, and each scalar potential can be decomposed into a discrete set of spherical models:
\begin{equation}
\phi_{lm}=R_l(r) P_l^{|m|}(\cos{\theta}) e^{i m \Phi}
\end{equation}
For a specific wavelength, because the dielectric constant is constant within each shell, $R_l(r)$ is a linear combination of the first and second kind of spherical Bessel functions within the two respective shells.
\begin{equation}
R_l(r) |_i = A_i j_l (k_i r) + B_i y_l (k_i r)
\end{equation}
We can solve for these coefficients with the transfer matrix of the interface.Thus we can calculate the transfer matrix of the whole system, by simply telescoping these solutions to individual interfaces
\begin{equation}
\begin{bmatrix}
A_{n+1} \\
B_{n+1}
\end{bmatrix}
=
M_{n+1,n}M_{n,n-1}...M_{3,2}M_{2,1}
\begin{bmatrix}
A_{1} \\
B_{1}
\end{bmatrix}
=
M
\begin{bmatrix}
A_{1} \\
B_{1}
\end{bmatrix}
\end{equation}
For the first shell, the contribution from the second kind of Bessel function must be zero because the second kind of Bessel function is singular at the origin. Thus, $A_1=1$, $B_1=0$. The coefficients of the surrounding layer are given by the transfer matrix element $A_{n+1}=M_{11}$ and $B_{n+1} = M_{21}$. To find the coefficients of this surrounding medium, we write the radical function as a linear combination of spherical Hankel functions:
\begin{equation}
R_l(r) |_{n+1} = C_{n+1} h_l^1 (k_{n+1}r) + D_{n+1} h_l^2 (k_{n+1} r)
\end{equation}
Here, $h_l^1(k_{n+1}r)$ and $h_l^2(k_{n+1}r)$ are the outgoing and incoming waves respectively, using the convention that fields vary in time as $e^{-i \omega t}$.
The reflection coefficients $r_l$ are given by:
\begin{equation}
r_l = \frac{C_{n+1}}{D_{n+1}} = \frac{M_{11}-i M_{21}}{M_{11} + i M_{21}}
\end{equation}

By solving for the reflection coefficients $r_l$, we can find the scattered power in each channel:
\begin{equation}
P_{l,m= \pm 1}^{sca} = \frac{\lambda^2}{16 \pi} (2 l + 1) I_0 |1-r_l|^2
\end{equation}
Lastly, by summing over all channels contributions of the TE and TM polarization (both of the $\sigma$ terms), we find the total scattering cross-section:
\begin{equation}
\sigma_{sca} = \sum\limits_{\sigma} \sum\limits_{l=1}^{\infty} \frac{\lambda^2}{8 \pi} (2 l + 1) |1 - r_{\sigma,l} | ^2
\end{equation}
For practical reasons, the $l$ summation did not go to $\infty$. Instead, before the training data was generated, the order of $l$ was slowly increased until the spectrum had converged and adding more orders would not change the result. For a typical calculation here, the order ranged from 4 $l$ terms, to 18 $l$ terms.

\subsection{Inverse Design with NN's}\label{inverseNN}

The arrangement of the network was a fully connected dense feedforward network. This smallest network we used had four layers, with 100 neurons per layer, which gave the network around 50,300 parameters. The network size was increased as the number of layers increased, with the maximum size being four layers with 300 neurons each for the particle with ten alternating layers. The input to this network was the thickness of each layer of the particle (with the fixed materials), and the output was the spectrum sampled at 200 points between 400 to 800 nanometers. We train the network using a batch size of 100, for around 16,000 epochs on most trials. The cost function that we use is the mean-square-error between each point on the spectrum and the 200 dimensional output of the neural network. This cost function was changed for the training versus the inverse design.

\begin{acknowledgments} This material is based upon work supported in part by the National Science Foundation under Grant No. CCF-1640012, as well as in part supported by the Semiconductor Research Corporation under Grant No. 2016-EP-2693-B. It is also supported in part by the U. S. Army Research Laboratory and the U. S. Army Research Office through the Institute for Soldier Nanotechnologies, under contract number W911NF-13-D-0001, as well as in part by the MRSEC Program of the National Science Foundation under award number DMR - 1419807. The authors furthermore thank Sam Peurifoy for reviewing and revising work. 

\end{acknowledgments}

\section{Supplementals}

\subsection{Details for the Comparison of Neural Networks with Inverse Design Algorithms}\label{inverseRuntime}




This section describes the results and details involved in comparing the inverse design runtimes. 

\begin{figure}[htbp]
\includegraphics[width=\linewidth]{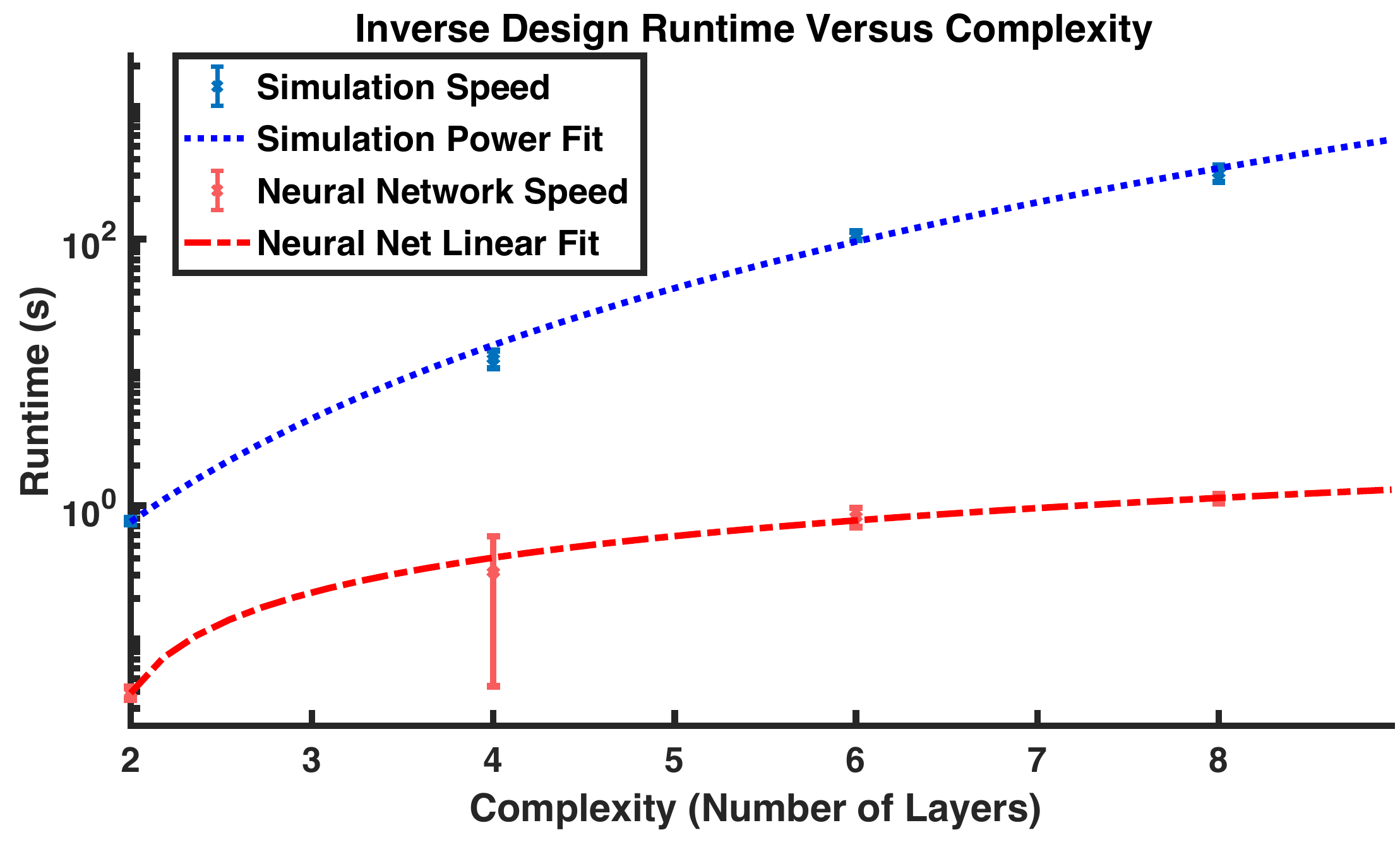}
\caption{Comparison of inverse design runtime versus complexity of the nanoparticle. The runtime of the numerical optimization is seen to increase more quickly than that of the neural network. The simulation is fit with a power fit (that finds an exponent of 4.5), and the neural network is fit with a linear fit.  
}
\label{fig:opt_graphs_1}
\end{figure} 

To compare the runtime of the neural network versus the numerical methods, we first had to train the networks to a given error threshold as described above. To allow for approximately the same error threshold even as the particles became more complex, the size of the Neural Network was increased as we considered more complex particles. The two layer particle had 30,000 parameters, while the four layer had 46,000 and the six layer had 151,000. Note that equivalent performance may possibly be achieved with much fewer parameters, as these architectures were not heavily optimized.

To establish a robust and comparable `accuracy cutoff' for the increasing complexity of the particles, we looked at the error rate of the numerical inverse design for the simulation. We did this because ultimately we wanted to perform a comparison of the neural network to the numerical inverse design on equal footing. Thus, we ensured that the neural network's accuracy cutoff during the training stage was below the error rate for the numerical inverse design. Effectively, we ran the numerical inverse design for five different particle configurations with the same number of layers, then found what the mean error rate of these tests were. This provided a robust and comparable `accuracy cutoff' that we could then use to figure out what the size of the neural network should be for each nanoparticle.

To get an equal footing comparison --- and trying to not bias our results to any particular choice of optimization method --- the comparison described here used the same inverse design optimization function for both the Neural Network's and the simulation. The approach described in the paper, reverse-backpropogation, gives comparable results; however it is difficult to do a fair comparison due to different mediums and different algorithms. Thus, after experimenting with several optimization functions, we used the same function for both the simulation and neural network, simply adding in the analytical gradient for the case of the neural network --- one of the key benefits.

The results of this are seen in Fig.~\ref{fig:opt_graphs_1}. From these results, it is evident that the runtime of the simulation for inverse design becomes large, while the neural network can handle more complex problems in the equivalent speed. The difference in the scaling --- the power fit for the simulation versus the linear fit of the neural network --- is one of the promising features about using this method.

\subsection{JAggregates}\label{jaggs}

Nanoparticles made from J-aggregates \cite{jAgg} often consist of a core shell geometry (normally made of a metal), surrounded by a layer of dielectric material, and then coated with a J-aggregate dye. This dye is peculiar because it couples with the metallic core to produce exciton resonance structures. These materials have allowed advanced studies of plasmon-exciton interactions and have been used to generate phenomena like induced transparency. These materials are powerful because the scientist can choose where these resonance structures and peaks happen in the material, and as such offer a level of customizability in designing nanoparticles. 

\begin{figure}[htb]
\includegraphics[width=\linewidth]{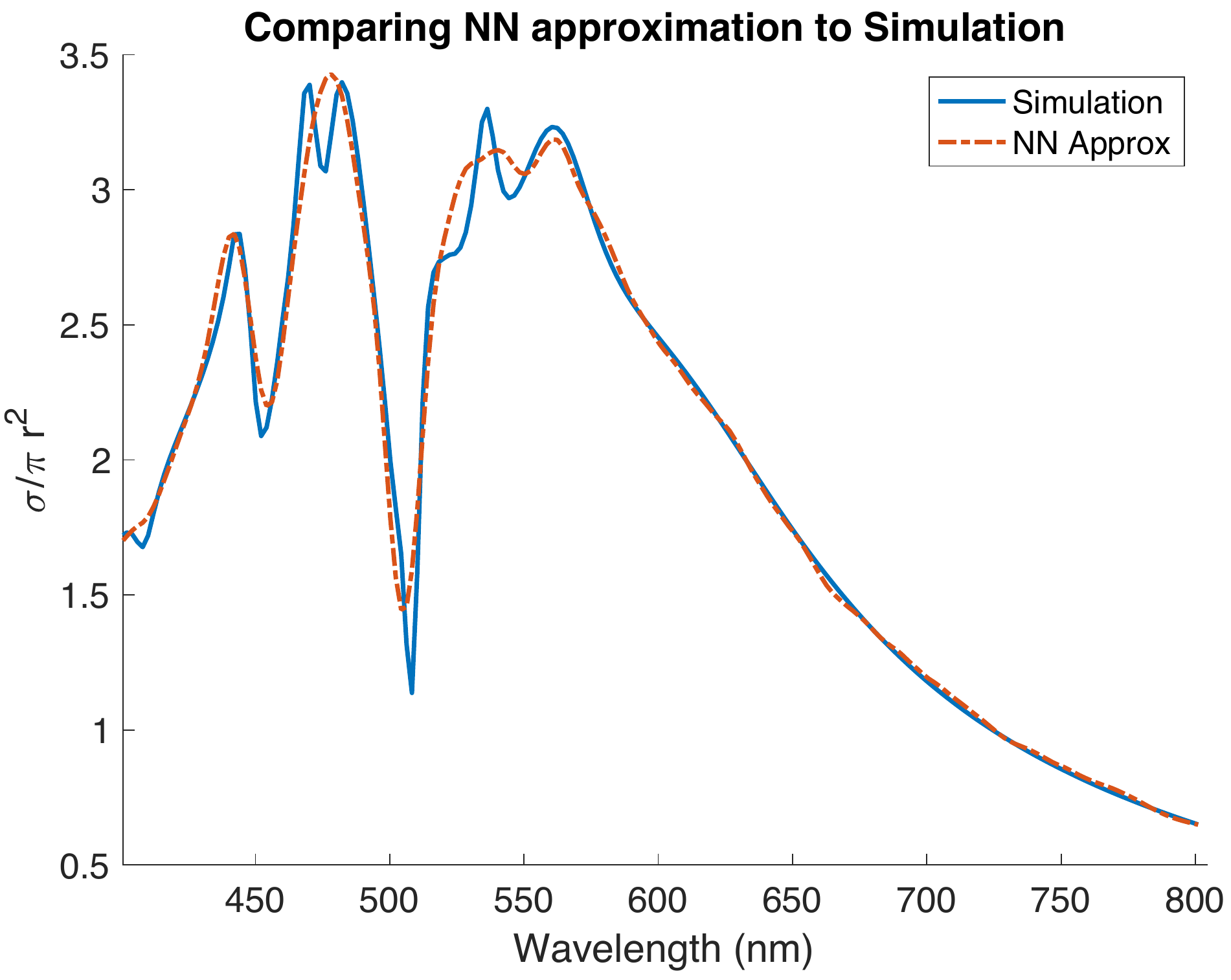}
\caption{Comparison of neural network approximation to the real spectrum for a particle made with a J-Aggregate material. The sharp peak in the spectrum is due to a resonance phenomenon in the J-Aggregate material, and can be customized for a variety of wavelengths. This result was generated from a particle not seen in the training data. 
\label{fig:jagg_comp}}
\end{figure}

\begin{figure}[htb]
\includegraphics[width=\linewidth]{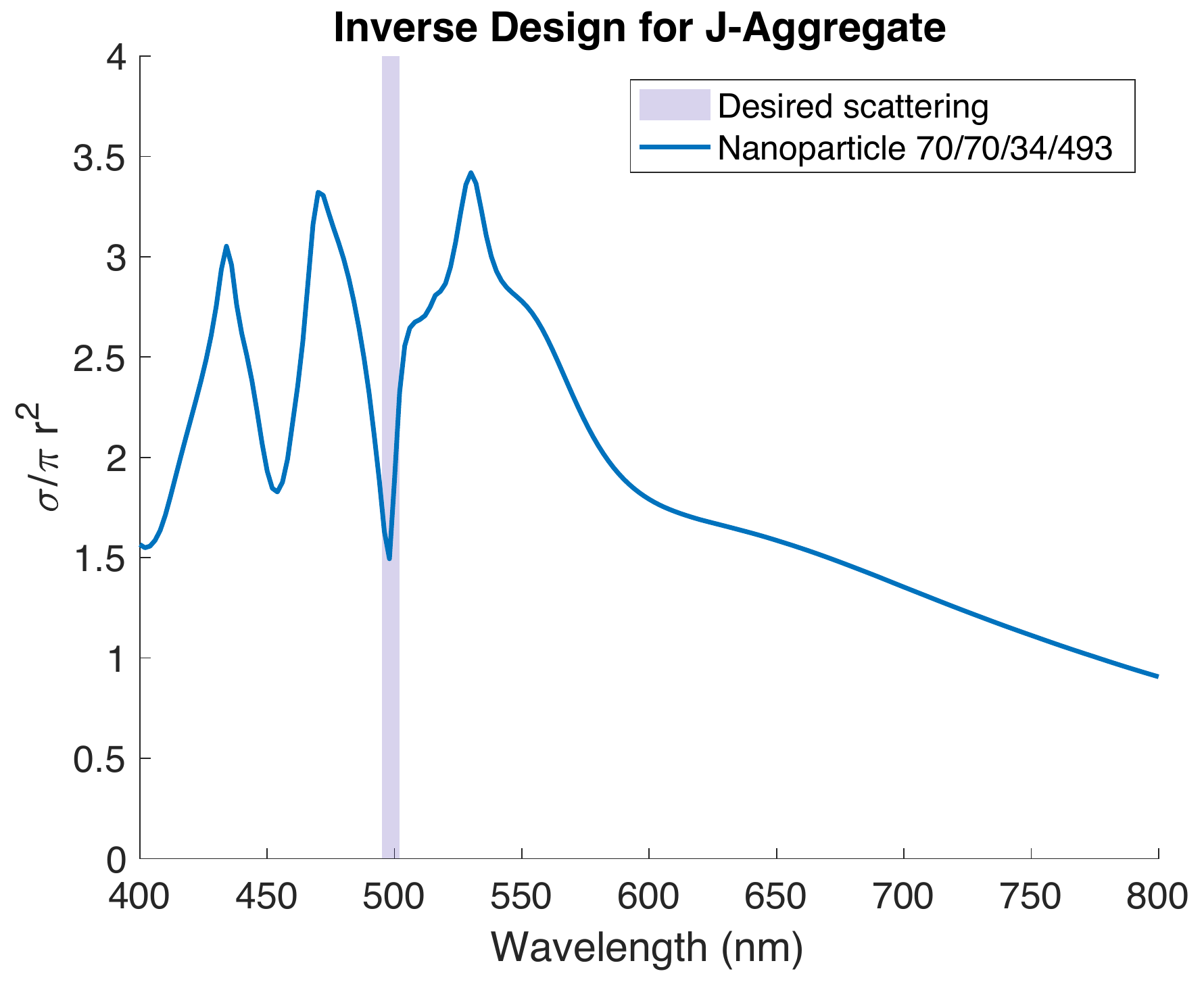}
\caption{Optimization of scattering at a particular wavelength using the J-Aggregate material. The sharp peaks in the result are possible due to the material properties of J-Aggregates, and present a complex scattering behavior.
\label{fig:jagg_desired}}
\end{figure}

This customizability --- which often can drastically change the spectra produced by these particles --- is a feature that makes these nanoparticles good litmus tests for the robustness of the network.

Here we considered a three layer particle made of a metallic silver core, a dielectric layer of silica, and an outside layer of J-aggregate dye. Each layer ranged from 30nm-70nm, and the J-aggregate dye had a different dielectric function for each training example. The frequency-dependent dielectric $\epsilon$ function of the J-aggregate dye was given by:

\begin{equation}
\epsilon(\omega) = \epsilon_0 + \frac{f \omega_0^2}{\omega_0^2-\omega^2-i \gamma \omega}
\end{equation}

Where $\epsilon_0 = 1.85$ , $f = 1.0$, $\gamma = .01$, $\omega = \frac{2 \pi}{\lambda}$. To control the value of this dielectric constant, $\lambda_{d}$ was varied in $\omega_0 = \frac{2 \pi}{\lambda_{d}}$ between 400 to 700nm. Note that this tunes the location of the resonance peak, but not the width of the peak. 

By varying the dielectric function, this means that each training example was different and had peaks located at significantly different locations. The inputs were only the thickness of each layer and the resonance peak of the J-aggregate material, no other information was supplied to the network.

Following the same procedure as above, the results of the network were visually inspected by testing a spectrum that had not been trained on --- Fig.\ref{fig:jagg_comp}. Results demonstrated that despite the spectra being different due to the changed resonance peak, the network was robust and could still approximate well. 

Similarly, we performed inverse design with J-aggregates --- Fig.\ref{fig:jagg_desired}. Due to the J-aggregate material, the sample space of spectra was much broader, and thus the results from the network were more attuned for the optimization. The sharper peaks allowed the network to find much more optimal configurations of the particle.

These results demonstrate that the network is robust even with sharp features in the spectrum, and furthermore that even with large sample spaces, the network is able to function as an optimization tool and create unique geometries.


\bibliography{PhotoNetPRL}

\end{document}